# Integrated Circuit Readout for the Silicon Sensor Test Station

*E. Atkin, A. Silaev, A. Kluev*
*MEPhi, Moscow*

*A. Voronin, M. Merkin, D. Karmanov, A. Fedenko*
*SINP MSU, Moscow*

Various chips for the silicon sensors measurements are described. These chips are based on 0.35 um and 0.18um CMOS technology. Several analog chips together with self-trigger /derandomizer one allow to measure silicon sensors designed for different purposes. Tracking systems, calorimeters, particle charge measurement system and other application sensors can be investigated by the integrated circuit readout with laser or radioactive sources. Also electrical parameters of silicon sensors can be studied by such test setup.

**Introduction**

Silicon sensors are using now in many HEP projects, cosmic rays experiments and nuclear physics. It requires different types of sensors: pad, microstrip with various geometries, thickness and other specifications.
On the other hand, for each project new readout is developed with rather complicate architecture, which not suitable for the test purpose. Mostly it touches R&D for new sensors. Also, to apply the readout is designed for one project very often impossible to do for next project sensor.
Test station for silicon sensors measurements must be flexible enough to provide quick adjustment the test station for new type of a sensor. We would note that using mass production components for the silicon sensors practically impossible and this is a reason why we cannot found any readout based on such components.
More than twenty years ago SINP MSU started the silicon sensor development. A lot of the sensors have been designed for many projects: HES/ZEUS (DESY), D0 (FERMILAB), ATIC, SVD-2 (IHEP), PHENIX(BNL) and other projects including R&D. Several years ago in collaboration with MEPhI we created a team for ASIC design. We started with simple chips using EUROPRACTICE as a way to get cheap software and production of small batch of chips. We choose a strategy that each chip might be used for two purposes: further development for more complicate schematics and using designed chip itself for some R&D and silicon sensors measurements.
The integrated circuit readout is a core of full test readout and requires additional components: analog and digital drivers, digital level convertors, ADCs, power supply system and components required for a specific task.

**Description of the chips readout**

The integrated circuit readout consists of four setups:
1. 16-channel charge sensitive amplifier (CSA) for single-sided silicon sensors, which are applied for systems with high dynamic range and two additional operational amplifiers (OP)
2. 8-channel CSA for double-sided silicon sensors for tracking system
3. 4-cnannel chip with Amplex structure and additional analog channels with high dynamic range
4. 4-channel comparator with derandomiser 4→2

**1. 16-channel charge sensitive amplifier** with switches feedback capacitors; which gives possibility to change the CSA gain. Additional OPs might be used for additional gain, shapers or track&hold circuitry.

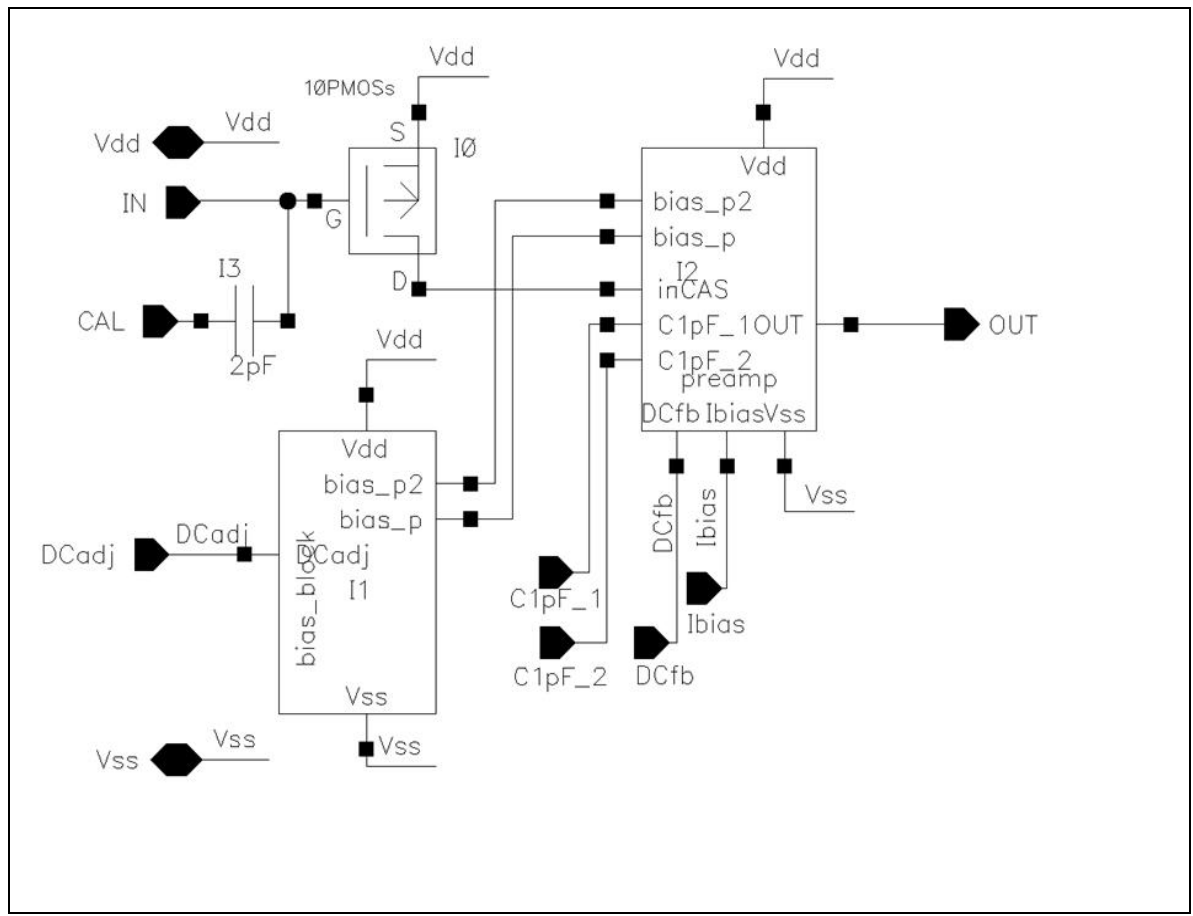

**Fig.1. Structure of the 16-channel CSA**

**2. 8-channel CSA** for double-sided silicon sensors also circuitry for silicon sensor dark current measurement up to 1 uA:

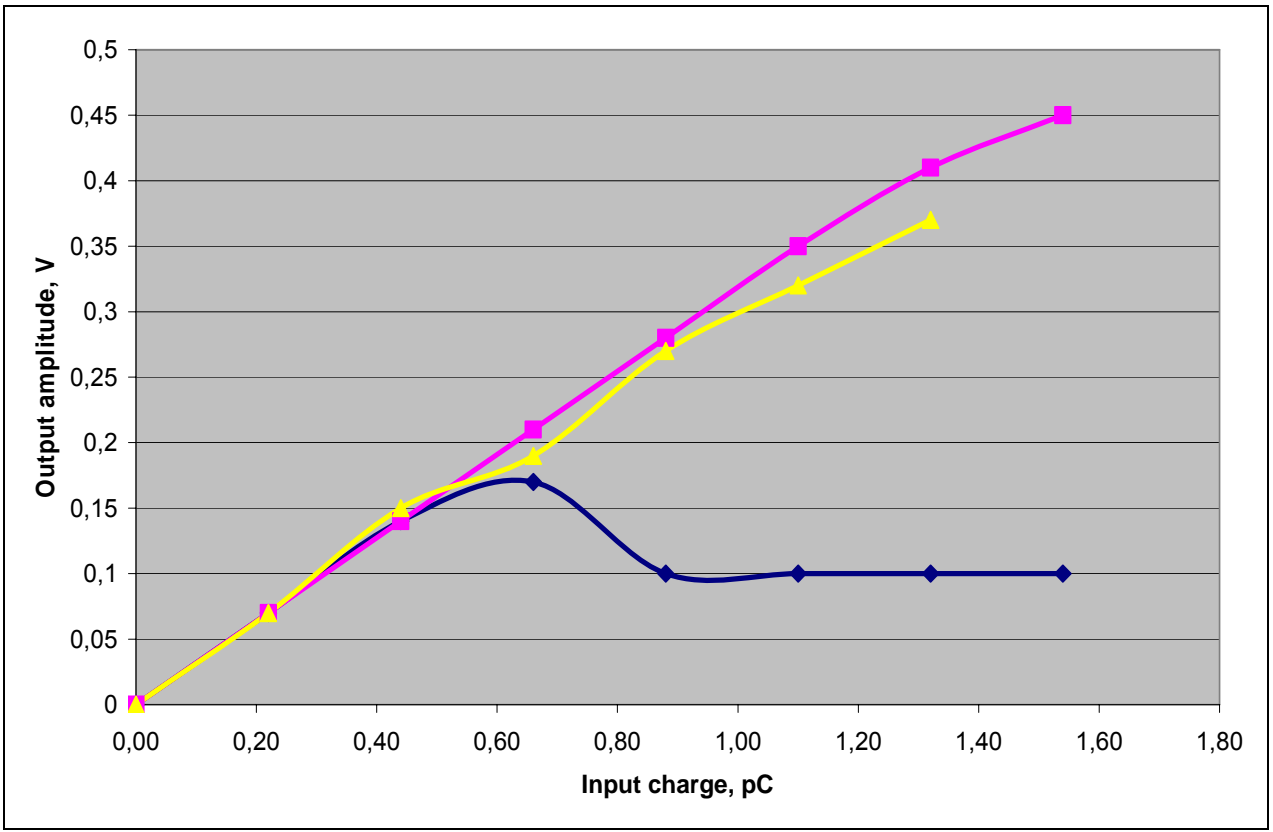

**Fig.2. Transfer function of the 8-channel CSA.** (Blue – n-side (Cd=0pF), Pink – p-side (Cd=0 pF), yellow - p-side (Cd=100 pF))

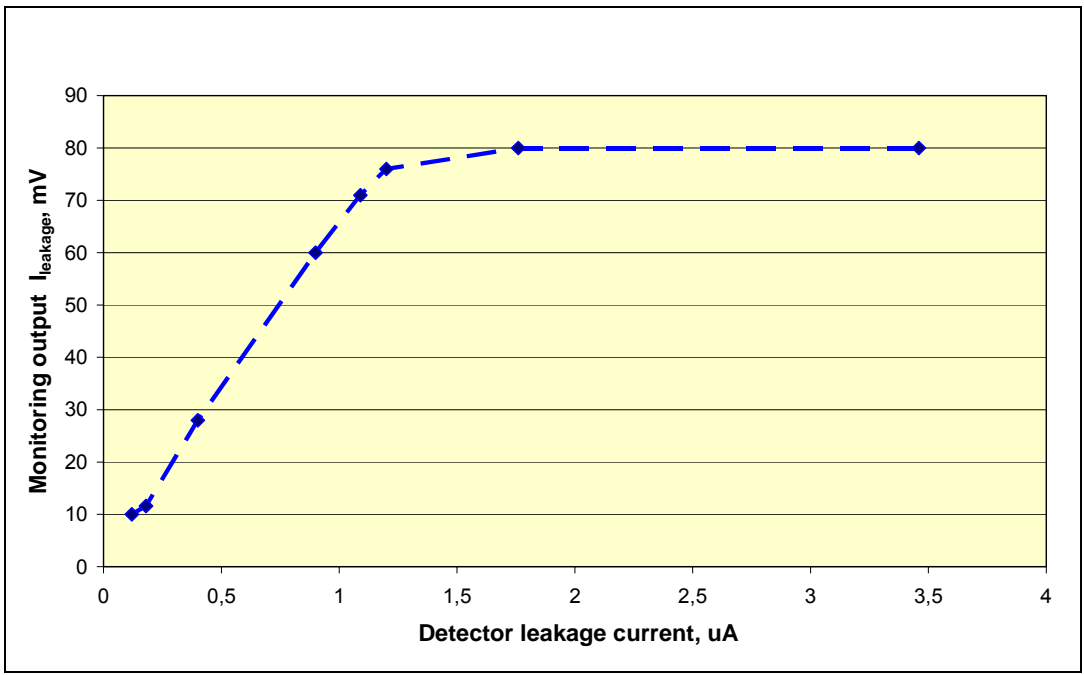

**Fig.3. Dark current data using the 8-channel CSA.**

**3. 4-cnannel chip with Amplex structure** and additional analog channels with high dynamic range contains CSAs, shapers, track and hold circuitry, analog multiplexor and output driver. A lot of additional adjustments are able to change parameters of chip. Additional dummy analog channels also contain CSAs, shapers, track&hold circuitry.

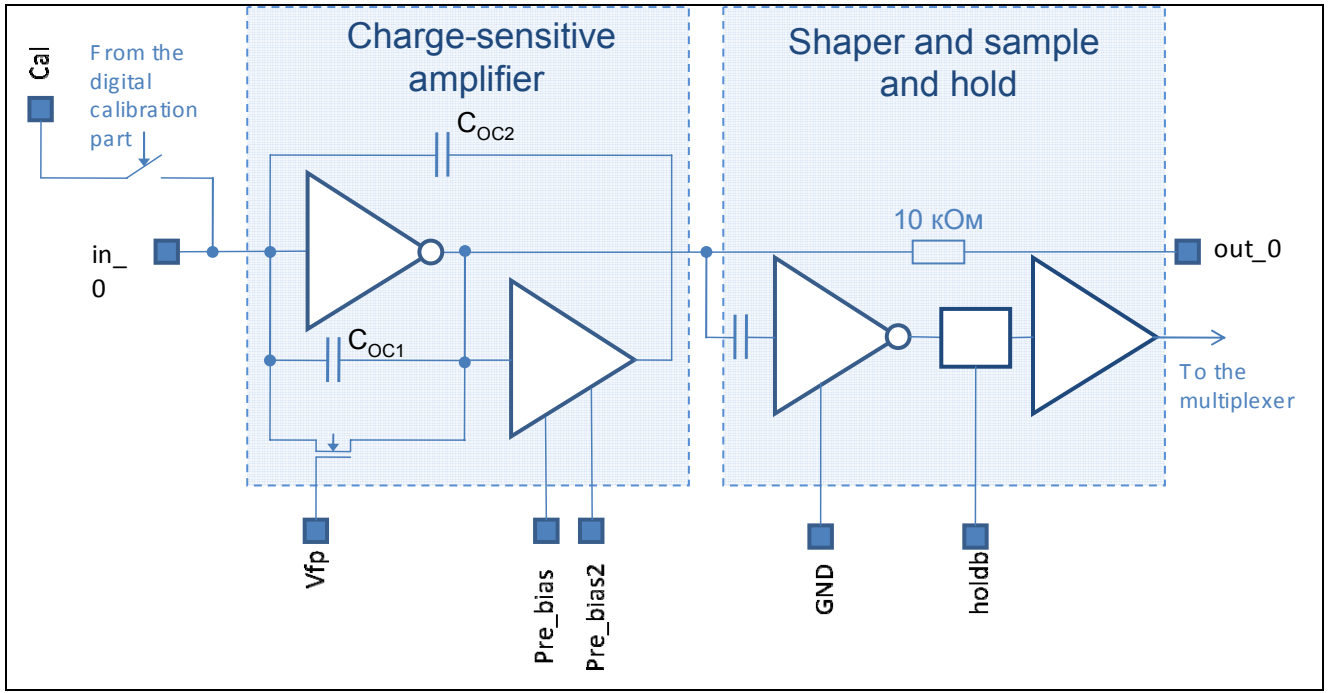

**Fig.4. Analog channel of the 4-channel chip with Amplex structure**

**4. 4-channel comparator with derandomiser 4→2** can be used as a self-trigger of a device or as a generator of the first level trigger for an external scheme. Also this chip contains structure which reduces number of ADC by half.

For these purposes two peak detectors are included and according their state "empty-busy" analog signals go to these peak detectors. Special arbitrage controller provides such functionality.

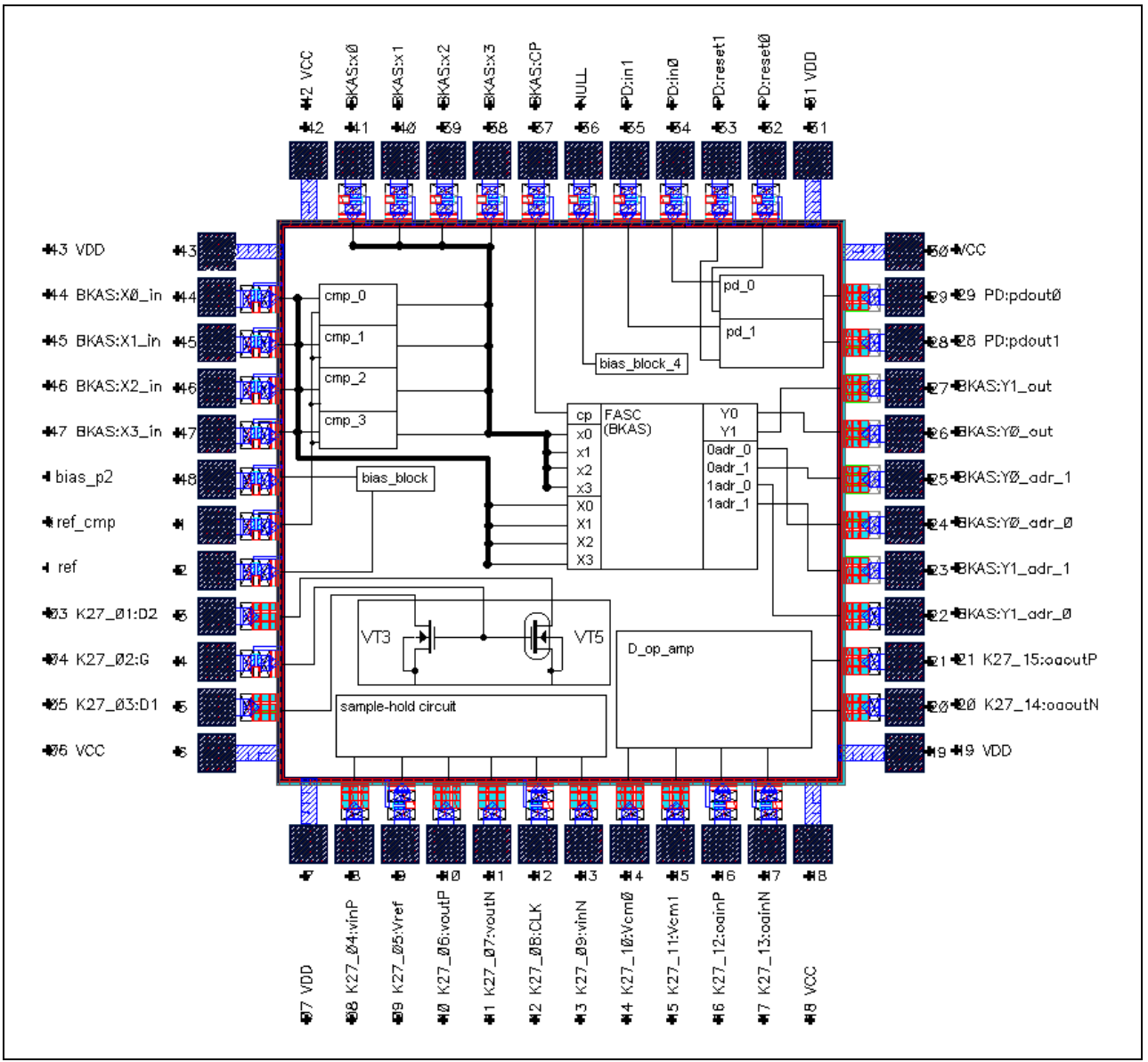

**Fig.5. 4-channel comparator with derandomiser 4→2 structure**

**Realization of the silicon sensor test readout using presented chips**

Silicon sensors couples inputs of CSA of any chips by a pitch adaptor:

1. 16-channel charge sensitive amplifier (CSA) for single sided silicon sensors which are applied for systems with high dynamic range and two additional operational amplifiers (OP)
2. 8-channel CSA for double sided silicon sensors for tracking system
3. 4-cnannel chip with Amplex structure and additional analog channels with high dynamic range

Each CSA has its own output pad, which can connect to 4-channel comparator and produce self-trigger or synchrosignal. To operate with signals on the output pads of each chip correctly, one needs to install some drivers, depending on the next equipment: it could be scope, ADC or other measurement units. We may adjust transistors DC biases of analog stages, gain, shaping time, because the biases have output pads and easily adjustable. Also one is able to observe signals from CSA, shapers and other scheme because it has output pads as well.

**Summary**

The silicon sensor test readout is realized by integrated circuit chip, which are designed by SINP MSU & MEPhI team. The test readout has been included into the test station for the silicon sensors. Parameters of the test readout can be easily adjusted for any type of a sensor.